\documentclass[reprint,pra,aps,nofootinbib,longbibliography,superscriptaddress]{revtex4-2}

\usepackage{amsmath,amssymb,amstext,amsthm} 
\usepackage[pdftex]{graphicx}
\usepackage{tabularx}
\usepackage{bm}
\usepackage{mathrsfs}
\usepackage{fancyhdr}
\usepackage{booktabs}
\usepackage{longtable}
\usepackage{multirow}
\usepackage{float}
\usepackage{scalerel}
\usepackage[normalem]{ulem}
\usepackage{dcolumn}
\usepackage{blindtext}
\usepackage{verbatim}
\usepackage{relsize}
\usepackage{musicography}
\usepackage{blindtext}
\usepackage{cancel}
\usepackage{physics}
\usepackage{epstopdf}
\usepackage{mathtools}
\usepackage{tensor}
\usepackage{color}
\usepackage[usenames,dvipsnames]{pstricks}
\usepackage{epsfig}
\usepackage{pst-grad} 
\usepackage{pst-plot} 
\usepackage{verbatim}
\usepackage{slashed}
\usepackage{dsfont}
\usepackage[english]{babel}

\usepackage[pdftex,pagebackref=false]{hyperref}
\hypersetup{
    unicode=false,          
    pdffitwindow=false,     
    pdfstartview={FitH},    
    pdfnewwindow=true,      
    colorlinks=true,        
    linkcolor=blue,         
    citecolor=green,        
    filecolor=magenta,      
    urlcolor=cyan           
}

\usepackage{braket}
\usepackage{bbold}
\usepackage{comment}

\allowdisplaybreaks

\newcommand{\mf}{\mathsf}

\newcommand{\ii}{\mathrm{i}}

\newcommand{\tc}[1]{\textsc{#1}}

\newcommand{\U}{\hat{U}}
\newcommand{\E}{\mathcal{E}}
\newcommand{\lettersec}[1]{\textbf{\textit{#1---}}}

\newtheorem*{proposition*}{Proposition}

\begin{document}

\title{A Universal Quantum Computer From Relativistic Motion}

\author{Philip A. LeMaitre}
\email{philip.lemaitre@uibk.ac.at}
\affiliation{University of Innsbruck, Institute for Theoretical Physics, Technikerstr. 21a, A-6020 Innsbruck, Austria}

\author{T. Rick Perche}
\email{trickperche@perimeterinstitute.ca}
\affiliation{Perimeter Institute for Theoretical Physics, Waterloo, Ontario, N2L 2Y5, Canada}
\affiliation{Department of Applied Mathematics, University of Waterloo, Waterloo, Ontario, N2L 3G1, Canada}
\affiliation{Institute for Quantum Computing, University of Waterloo, Waterloo, Ontario, N2L 3G1, Canada}

\author{Marius Krumm}
\affiliation{University of Innsbruck, Institute for Theoretical Physics, Technikerstr. 21a, A-6020 Innsbruck, Austria}

\author{Hans J. Briegel}
\affiliation{University of Innsbruck, Institute for Theoretical Physics, Technikerstr. 21a, A-6020 Innsbruck, Austria}

\date{\today}

\begin{abstract}

We present an explicit construction of a relativistic quantum computing architecture using a variational quantum circuit approach that is shown to allow for universal quantum computing. The variational quantum circuit consists of tunable single-qubit rotations and entangling gates that are implemented successively. The single qubit rotations are parameterized by the proper time intervals of the qubits' trajectories and can be tuned by varying their relativistic motion in spacetime. The entangling layer is mediated by a relativistic quantum field instead of through direct coupling between the qubits. Within this setting, we give a prescription for how to use quantum field-mediated entanglement and manipulation of the relativistic motion of qubits to obtain a universal gate set, for which compact non-perturbative expressions that are valid for general spacetimes are also obtained. We also derive a lower bound on the channel fidelity that shows the existence of parameter regimes in which all entangling operations are effectively unitary, despite the noise generated from the presence of a mediating quantum field. Finally, we consider an explicit implementation of the quantum Fourier transform with relativistic qubits.

\end{abstract}

\maketitle

\lettersec{Introduction} Over the past few decades, traditional notions of computation have been transformed by important developments in quantum information and quantum computation~\cite{nielsen2010quantum}, with attention almost exclusively directed toward non-relativistic setups. However, with the rising interest in distributed computing~\cite{DistributedQC} and space-based quantum networks using satellites~\cite{QKDSatellite}, it is becoming increasingly relevant to study and understand the relativistic regime. Along this line, the field of relativistic quantum information (RQI) has developed numerous protocols that rely on the relativistic features of spacetime and quantum fields, such as entanglement harvesting~\cite{Valentini1991,Reznik2003,Pozas-Kerstjens_Martín-Martínez_2015,Pozas-Kerstjens_Martín-Martínez_2016,Mendez-Avalos_Henderson_Gallock-Yoshimura_Mann_2022,relHarvesting} and quantum energy teleportation~\cite{hotta2008quantum,HottaEntanglement,HottaDistance,teleportExperiment}. It is then natural to ask how these relativistic features could be incorporated into the design and operation of quantum computers, and what the implications of such a distributed setting would be.

Previous works regarding relativistic quantum computing~\cite{Martin-Martinez_Aasen_Kempf_2013,Bruschi_Dragan_Lee_Fuentes_Louko_2013,Martín-Martínez_Sutherland_2014,Bruschi_Sabín_Kok_Johansson_Delsing_Fuentes_2016,Layden_Martin-Martinez_Kempf_2016,AsplingLawler2024} have shown that approximate one and two-qubit quantum gates can be constructed through control of the relativistic motion of qubits which interact with a quantum field using the Unruh-DeWitt (UDW) model~\cite{PhysRevD.14.870,Hawking:1979ig}, suggesting that one should be able to construct a universal $N$-qubit relativistic quantum computer. However, in fundamental descriptions, all interactions are mediated by quantum fields that give rise to noisy operations~\cite{LoukoCurvedSpacetimes,Pozas-Kerstjens_Martín-Martínez_2015} and make it unclear whether the gates implemented in this manner are viable for effective quantum computation. Moreover, explicit configurations beyond 2 qubits have not yet been discussed, which is problematic due to the difficulty in controlling the transmission of information through quantum fields, making claims of universal quantum computing hard to follow. Both controllability of the field-mediated entanglement and quantification of the decoherence due to the interaction with the field are essential for the implementation of faithful quantum operations. Due to these issues, no theoretical implementation of a practical quantum algorithm has been achieved utilizing UDW qubits coupled to quantum fields.

A quantum computing architecture that can naturally incorporate UDW time-evolution into its structure, and which would allow for systematic manipulation of the qubits' spacetime trajectories while also granting us the flexibility with these parameters to achieve universal quantum computation, is that of a \emph{variational quantum circuit} (VQC)~\cite{biamonte2021universal}. VQCs have become increasingly popular in the last several years as the go-to implementation of quantum machine learning (QML) ~\cite{Cerezo_Arrasmith_Babbush_Benjamin_Endo_Fujii_McClean_Mitarai_Yuan_Cincio_2021,Leone_Oliviero_Cincio_Cerezo_2024}. This popularity has led to many results with both hardware and software implementations of VQCs, providing a convenient vehicle to implement relativistic quantum computing. 

In this manuscript, we present a universal quantum computer with non-perturbative expressions for its universal gate set in a new regime, valid for general spacetimes, which is comprised of $N$ relativistic UDW qubits distributed through spacetime resembling the so-called hardware-efficient configuration of a VQC~\cite{Cerezo_Arrasmith_Babbush_Benjamin_Endo_Fujii_McClean_Mitarai_Yuan_Cincio_2021,Leone_Oliviero_Cincio_Cerezo_2024}. The trainable parameters of the VQC then become the parameters defining the qubits' trajectories in spacetime. We further demonstrate the application of our platform by implementing a standard quantum algorithm: the quantum Fourier transform. 

\lettersec{Variational Quantum Computation} Current quantum computers typically implement a limited number and type of operations effectively due to experimental restrictions and noise. This scenario is not so different when one considers fundamental interactions with a quantum field, which limit the operations that can be directly implemented. It is then important to build unitaries that represent relevant quantum circuits from a limited set of quantum operations~\cite{Cerezo_Arrasmith_Babbush_Benjamin_Endo_Fujii_McClean_Mitarai_Yuan_Cincio_2021,Leone_Oliviero_Cincio_Cerezo_2024}. 

In a typical implementation of a VQC, one considers a parametrized unitary $\hat{U}_{\bm \theta}$ that depends on a set of \emph{trainable parameters} $\bm \theta$, which are chosen through the process of minimizing a loss function $\mathcal{L}(\bm\theta)$ representing a specific goal. Successful minimization of $\mathcal{L}(\bm \theta)$ results in a set of parameters $\bm\theta$ such that $\hat{U}_{\bm \theta}$ performs the desired computation, represented by a target unitary $\hat{U}$. The loss function usually depends on the trainable parameters $\bm \theta$, a set of input states \(\{\rho_k\}\), and a set of observables \(\{\hat{\mathcal{O}}_k\}\); common expressions for the loss only depend on the expectation values $E_k(\bm \theta) = \text{Tr}\left(\hat{\mathcal{O}}_k \hat{U}_{\bm \theta} \hat{\rho}_k \hat{U}^\dagger_{\bm \theta}\right)$ and have the form $\mathcal{L}(\bm \theta) = \sum_k f_k\left(E_k(\bm \theta)\right)$ , for some set of functions \(\{f_k\}\). The minimum of $\mathcal{L}(\bm \theta)$ then corresponds to the case $\hat{U}_{\bm \theta} = \hat{U}$, for example.

For the choice of $\hat{U}_{\bm \theta}$, most contemporary literature focuses on \emph{hardware-efficient} assumptions, which use a layered circuit ansatz $\hat{U}_{\bm \theta } = \prod_{\ell = 1}^D \U^{(\ell)}_{\bm \theta}$, where $D$ is the depth of the circuit and each layer decomposes into two steps via $\U^{(\ell)}_{\bm \theta} = \U_{\mathrm{ent}} \U_{\mathrm{para}}(\bm \theta^{(\ell)})$. Usually, the parametrized part $\U_{\mathrm{para}}(\bm \theta^{(\ell)})$ consists of parallel arbitrary single-qubit rotations \cite{Cerezo_Arrasmith_Babbush_Benjamin_Endo_Fujii_McClean_Mitarai_Yuan_Cincio_2021,Leone_Oliviero_Cincio_Cerezo_2024}. For a circuit involving $N$ qubits, the trainable parameters on each layer can be written as $\bm \theta^{(\ell)} =(\theta_{1}^{(\ell)},\vartheta_{1}^{(\ell)},\varphi_{1}^{(\ell)},...,\theta_{N}^{(\ell)},\vartheta_{N}^{(\ell)},\varphi_{N}^{(\ell)})$, where $\ell$ is the layer and the three different angles define the angle and axis of rotation. In this setup, a typical parametrization can be written as
\begin{equation} \label{Equation:UDWRotate}
    \hat{U}_{\text{para}}(\bm \theta^{(\ell)})  = \prod_{i=1}^N \hat{U}_{\bm n_i}({\theta}^{(\ell)}_i)= \prod_{i=1}^N e^{- \ii \theta_{i}^{(\ell)} \bm n(\vartheta_{i}^{(\ell)},\varphi_{i}^{(\ell)}) \cdot \hat{\bm \sigma}_i}\,,
\end{equation} 
where $\hat{\bm \sigma}_i$ denotes the vector of Pauli matrices acting on qubit $i$ and $\bm n_i=\bm n(\vartheta_i^{(\ell)},\varphi_i^{(\ell)}) = (\sin\vartheta_i^{(\ell)}\cos\varphi_i^{(\ell)},\sin\vartheta_i^{(\ell)}\sin\varphi_i^{(\ell)},\cos\vartheta_i^{(\ell)})$ specifies the rotation direction. The entangling unitary $\U_{\mathrm{ent}}$ typically consists of products of 2-qubit gates, which entangle the corresponding qubit pairs. 

Due to hardware restrictions, current quantum computers can usually only implement a limited set of particular entangling gates, such as $\mathrm{CNOT}$ or $\mathrm{CZ}$ gates, to specific qubit configurations (e.g. nearest neighbors). Despite this restricted connectivity, by picking adequate values for the parameters $\bm \theta^{(\ell)}$ at each layer and ensuring there are sufficient layers, it is possible to obtain a wide class of $N$-qubit unitaries. Moreover, adequate choices of $\hat{U}_{\mathrm{ent}}$ allow for universal quantum computation~\cite{biamonte2021universal,Leone_Oliviero_Cincio_Cerezo_2024}.

\lettersec{A Relativistic Model} Our goal is to use UDW qubits~\cite{PhysRevD.14.870,Hawking:1979ig} to construct universal gate sets from relativistic motion, consisting of arbitrary single qubit rotations and entangling gates that are implemented through an interaction with a quantum field. We then show how to construct a relativistic VQC (RVQC) from this gate set.

We consider $N$ qubits that move along timelike trajectories in a globally hyperbolic spacetime $\mathcal{M}$ that admits a global timelike coordinate $t$. The interaction between the qubits is mediated by a scalar quantum field $\hat{\phi}(\mf x)$, and their local dynamics depend on their specific trajectories in spacetime.  

\textit{The Local Operations---} We start by describing how to implement the single qubit rotations in a relativistic setup. At each step where the local operations are applied, we consider that each of the qubits follows a classical trajectory $z^\mu_i(\tau_i) = (t(\tau_i),\bm x(\tau_i))$ in spacetime for a time $t_0<t(\tau_i)<t_1$, so that the total time duration of this evolution in the coordinate time $t$ is $\Delta t = t_1 - t_0$. Throughout their motion, the qubits' internal dynamics are prescribed by their free Hamiltonians, described in their rest frames as
\begin{equation}
    \hat{H}_i = \frac{\Omega_i}{2} \bm n(\vartheta_i,\varphi_i) \cdot \hat{\bm \sigma}_i,
\end{equation}
where $\bm n_i = \bm n(\vartheta_i,\varphi_i)$ is a unit vector defined by the azimuthal angle $\vartheta_i$ and polar angle $\varphi_i$; $\Omega_i$ corresponds to each qubit's energy gap. Over a proper time interval $\Delta \tau_i$, the local unitary implemented by the internal dynamics of each qubit is a Bloch rotation around the $\bm n_i$ axis, which takes the form of each factor in Eq.~\eqref{Equation:UDWRotate} with parameters $\theta_i = \Omega_i\Delta\tau_i/2$. Identifying the set $\bm \theta = \{\theta_i, \vartheta_i, \varphi_i\}_i$ as our trainable parameters and defining the product of these local unitaries as a trainable layer, we get the first part of a VQC~\cite{Cerezo_Arrasmith_Babbush_Benjamin_Endo_Fujii_McClean_Mitarai_Yuan_Cincio_2021}. Importantly, the $\bm \theta$ depend not only on the internal energy gap of the qubits $\Omega_i$, but also on the proper time difference experienced by each qubit while undergoing its trajectory. For instance, if each qubit undergoes a circular trajectory of radius $R$ with angular frequency $\omega_i$ in Minkowski spacetime, $\Delta \tau_i = \sqrt{1 - R^2\omega_i^2}\Delta t$, so that $\theta_i$ is controlled by changing the motion of the qubit in spacetime, rather than by controlling its energy gap. In essence, the details of each trajectory affect the rotation angle applied to each qubit~\cite{footnote1}.

\textit{The Entangling Operations---} The entangling channel $\mathcal{E}$ is implemented by allowing the qubits to interact with a quantum field via the UDW  model. The interaction of a two-level UDW qubit with a scalar quantum field $\hat{\phi}(\mf x)$ is often prescribed by the interaction Hamiltonian density~\cite{us,us2,Landulfo,Perche_2024}
\begin{equation}
    \hat{h}_i(\mf x)  = \lambda \Lambda_i(\mf x) \hat{\mu}_i \hat{\phi}(\mf x),
\end{equation}
where $\lambda$ is a coupling constant, $\hat{\mu}_i$ is an operator acting in the Hilbert space of qubit $i$ and $\Lambda_i(\mf x)$ is a spacetime smearing function, defining the shape of the interaction in spacetime. The function $\Lambda_i(\mf x)$ is strongly supported around the qubit's trajectory, $\mf z_i(\tau_i)$ and its spatial extent corresponds to the physical realization of the qubit. In our setup, we consider that the support of $\Lambda_i(\mf x)$ is effectively contained in the region $t_1 < t < t_2 = t_1 + \Delta t$, so that the interaction takes place after the local unitaries Eq.~\eqref{Equation:UDWRotate} are applied to the qubits. For simplicity, we assume that while the interactions with the field take place the qubit's internal dynamics are switched off (i.e. $\Omega_i = 0$), so that their evolution is entirely determined by their interaction with the field.

The dynamics of the qubits can be computed by applying the time evolution operator associated with the total interaction Hamiltonian density, $\sum_{i=1}^N \hat{h}_i(\mf x)$, and tracing out the field. This results in a quantum channel $\mathcal{E}$, which can entangle the qubits with each other (for $\hat{\mu}_i \neq \openone$), but also with the field. For simplicity, we assume the field to be in a quasi-free state~\cite{footnote2} $\hat{\rho}_\phi$ and we compute the channel $\mathcal{E}$ explicitly in Appendix~\ref{Appendix:Calculations}, where we show that it can be decomposed as
\begin{equation}
    \E(\hat{\rho}_0) = \hat{U}_\tc{c} \E_\phi(\hat{\rho}_0) \hat{U}_\tc{c}^\dagger\,,
\end{equation}
where $\E_\phi$ is a quantum channel that introduces decoherence due to entanglement between the qubits and the field and $\hat{U}_\tc{c}$ is a unitary that entangles different qubits, given by
\begin{equation} \label{Equation:UDWEntangle}
    \hat{U}_\tc{c} = e^{- \frac{\ii\lambda^2}{2}\sum_{{i<j}} \Delta_{ij} \hat{\mu}_i \hat{\mu}_j}\,,
\end{equation}
where $\Delta_{ij}$ are parameters that depend on retarded propagation of the qubit interaction regions through the field. These usually behave proportionately to the quotient of the interaction time of the qubits with the field and their spatial separation. A key assumption for the entangling unitary to act significantly between the qubit pairs is that $|\lambda^2 \Delta_{ij}| \sim 1$. This can be accomplished whenever the qubits' interaction regions are in causal contact for enough time, which allows the field to generate a sufficient amount of entanglement between the qubits. This couples all qubits simultaneously to the field, yielding a globally entangling unitary on all $N$ qubits. 

The noise introduced by the decohering channel $\mathcal{E}_\phi$ is often small under the assumption that $\lambda$ is sufficiently small. In Appendix~\ref{Appendix:Fidelity} we show that for any input state $\hat{\rho}_0$ the fidelity between $\mathcal{E}(\hat{\rho}_0)$ and $\hat{U}_\tc{c}\hat{\rho}_0\hat{U}_\tc{c}^\dagger$ has the lower bound
\begin{equation}
    F\left(\mathcal{E}(\hat{\rho}_0),\hat{U}_\tc{c}\hat{\rho}_0\hat{U}_\tc{c}^\dagger\right) \geq e^{-2\lambda^2N^2 \mathcal{W}}\Tr(\hat{\rho}_0^2)\,, \label{Equation:FidelityExponentialBoundText}
\end{equation}
where $\mathcal{W}$ is a field-dependent parameter that determines the local noise experienced by each qubit; it is often of order unity when the field is in a state that has a low particle number, as perceived by the qubits~\cite{footnote3}. Notice that while the term above is exponentially decaying with the number of qubits, the assumption that $\lambda^2 N^2 \mathcal{W}\ll 1$ guarantees that this noise can be neglected~\cite{footnote4}. In this case, the channel $\mathcal{E}_\phi$ acts as a weakly decohering channel. Still, the small field interaction noise is always present regardless of the specifics of the interaction, since it is present even when the field is in its vacuum state. 

Combining the assumption required for the entangling unitaries to act significantly on the qubits ($|\lambda^2 \Delta_{ij}| \sim 1$) with the assumption required for the noise experienced by the qubits to be sufficiently small ($\lambda^2 N^2 \mathcal{W} \ll 1$), we find that the condition for this setup to provide a suitable entangling gate is $N^2  \mathcal{W} \ll |\Delta_{ij}|$. Considering that $\Delta_{ij}$ usually scales with the interaction times~\cite{Perche_2024}, one simply requires that the setup couples a number of qubits ($N_q \leq N$) that is sufficiently smaller than the interaction time divided by the qubit separation. 

When one considers successive applications of the entangling channel $\mathcal E$, an additional assumption is required to ensure that the initial state of the field can be treated as quasi-free~\cite{Haag}. After one application of the channel $\mathcal{E}$, the state of the field also evolves due to the interaction with the qubits, and in general the field's final state will not be quasi-free. However, field excitations produced by the qubits dissipate away at the speed of light. Under the assumption that the time required to implement the free unitaries (while the qubits are not coupled to the field) is large compared to the qubits' separation, one can effectively treat the local state of the field probed by the qubits as its previous quasi-free initial state.

\begin{figure}[h!]
    \includegraphics[width = 1 \linewidth]{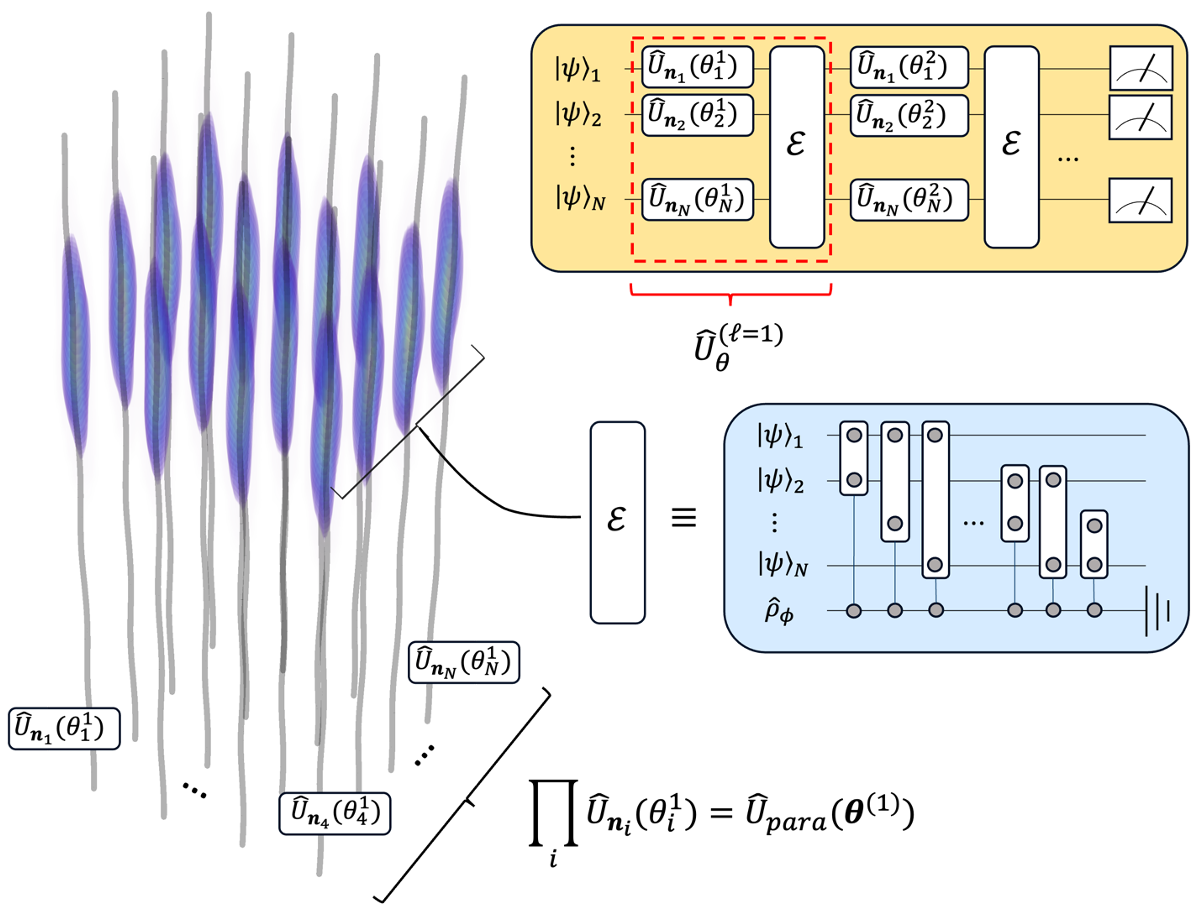}
    \caption{The RVQC model. On the left we have an array of UDW qubits that each follow a worldline through spacetime. During the pictured time interval of the worldlines, all qubit trajectories are manipulated to set the qubit rotation angles. They then each interact with a scalar quantum field in the purple spacetime regions becoming entangled; this corresponds to applying one layer $\hat{U}_{\bm \theta}^{(\ell)}$ of the RVQC, which is shown in the lower blue circuit diagram. Layers are repeatedly applied and the output is measured after the final layer is applied, as in the top golden circuit diagram.}
    \label{Figure:RVQC}
\end{figure}

\textit{The RVQC---} A relativistic analogue of the hardware-efficient ansatz (Fig. \ref{Figure:RVQC}) can be assembled from $D$ products of the local unitaries $\U_\text{para}(\bm \theta)$ and the entangling channel $\mathcal{E}(\hat{\rho}_0) \approx \hat{U}_\tc{c} \hat{\rho}_0\hat{U}_\tc{c}^\dagger$:
\begin{align} \label{Equation:RVQC}
\hat{U}_{\bm{\theta}} = \prod_{\ell=1}^{D} \hat{U}_{\bm \theta}^{(\ell)}\,= \prod_{\ell=1}^{D} \hat{U}_\tc{c}\hat{U}_\text{para}(\bm{\theta}^{(\ell)})\,.
\end{align}
To extract the result of applying  $\hat{U}_{\bm{\theta}}$ to the initial state $\hat{\rho}_0$, one computes the expected values of certain observables $\hat{\mathcal{O}}_k$ (e.g. Pauli strings). In the QML setting, the above would constitute one round of training and there would be the additional step of running the parameter optimization routine with respect to a given loss function $\mathcal{L}(\bm \theta)$~\cite{Cerezo_Arrasmith_Babbush_Benjamin_Endo_Fujii_McClean_Mitarai_Yuan_Cincio_2021}. 

Our RVQC possesses arbitrary 1-qubit rotations and an entangling gate, but it is still unclear if it represents a universal quantum computer --- that is, if it is capable of approximating any unitary operator arbitrarily well~\cite{Deutsch_Barenco_Ekert_1995}. The main obstacle to universality comes from the field interactions spreading entanglement out in all directions, resulting in a fully connected qubit configuration that could be too entangled to do useful computation. However, in Appendix~\ref{Appendix:Universality} we show that it is possible to efficiently compile entangling gates that act only on single qubit pairs by combining $N$ successive applications of the global entangling operator $\hat{U}_\tc{c}$ with local operations, proving that our RVQC is indeed a universal quantum computer. Compilation of a universal gate set for running arbitrary, fixed quantum circuits with our framework is also possible using the same methodology.

\lettersec{An Explicit Setup} We now consider an explicit setup in Minkowski spacetime where the qubits undergo circular motion while the local unitaries are applied and are inertial while interacting with a massless scalar field. We use an inertial time coordinate $t$ (lab coordinate time) and consider that the qubits undergo circular motion of radius $R$ and frequency $\omega_i$ while not interacting with the field. We also assume that their energy gap is the same, $\Omega_i = \Omega$. In this case, the trainable parameters become $\theta_i = \Omega \sqrt{1 - R^2 \omega_i^2}\Delta t/2$, where $\Delta t$ is the time of the qubits' free dynamics, as measured in the inertial time coordinate $t$. The training parameters $\vartheta_i$ and $\varphi_i$ still control the direction and magnitudes of the single-qubit rotations implemented during the qubits' free evolution. 

After their free evolution, the qubits return to a state of motion at rest with respect to the lab frame undergoing trajectories $z^\mu_i(t) = (t,\bm x_i)$, where $\bm x_i$ denotes their spatial coordinates. For convenience we define \mbox{$L_{ij} = |\bm x_i - \bm x_j|$}. We assume that the field is in the vacuum state and that the qubits' interaction with the field takes the shape of spacetime Gaussians
\begin{equation}
    \Lambda_i(\mf x) = e^{- \frac{(t - t_\tc{c})^2}{2T^2}}\frac{e^{- \frac{|\bm x - \bm x_i|^2}{2\sigma^2}}}{(2\pi \sigma^2)^\frac{3}{2}} \label{Equation:GaussianSmearing}
\end{equation}
where $T$ controls the interaction duration, $\sigma$ is the effective extension of the qubits, and $t_\tc{c}$ is the center of the interaction duration in time. We assume that the interaction lasts for $t_1<t<t_1 + 14T$, with $t_\tc{c} = 7T$. This ensures that less than a billionth of the support of $\Lambda_i(\mf x)$ is outside the time interval at which the entangling gate is applied to the qubits, and one can use asymptotic results~\cite{Perche_2024} for the computation of the channel $\mathcal{E}$. 

In this setup, it is possible to find asymptotic results for all parameters that determine the channel $\mathcal{E}$ (see Appendix~\ref{Appendix:ClosedForm}). Under the assumption that $\sigma\ll L_{ij} \ll T$,~\cite{footnote5} we find that the parameters $\Delta_{ij}$ take the form
\begin{equation}
    \Delta_{ij} = - \frac{1}{2\sqrt{\pi}}\frac{T}{L_{ij}},
\end{equation}
so that the phases that determine the entangling unitary $\hat{U}_\tc{c}$ are inversely proportional to the spatial separation between the qubits.

This explicit setup is analytically tractable and gives a simple form for the lower bound on the fidelity. For a pure input state $\ket{\psi_0}$, we find that
\begin{equation}
    F\left(\mathcal{E}(\ket{\psi_0}\!\!\bra{\psi_0}),\hat{U}_\tc{c}\ket{\psi_0}\right) \geq e^{- \frac{\lambda^2 N^2}{2\pi}}.
\end{equation}
For instance, with a coupling constant $\lambda$ of the order of $10^{-4}$, one could have $N=100$ qubits, and still keep a fidelity of $99.998\%$ for each layer. Concatenating many gates and using these estimates, one would reach $D=100$ layers while maintaining a lower bound for the fidelity of $99.8\%$. Although we are neglecting other sources of noise that could affect the qubits during the interaction with the field, our point is that the noise introduced by the field is negligible.

As a simple illustration of the RVQC, we consider the quantum Fourier transform for $N=6$ qubits. We fix the qubit geometry by arranging them into a $3\times 2$ square lattice with separations of \mbox{$(\Delta x, \Delta y) = T\lambda^2(5, 3)$}, where $T=10^8$ and $\lambda = 10^{-4}$. Then we set the spatial smearing width to $\sigma = 10^{-8}$ and fix the entangling gate operators $\hat{\mu}_i = \hat{\sigma}_z$ for each qubit. We consider $D=50$ layers and find that we are able to approximate the quantum Fourier transform $\hat{U}_{\text{QFT}}$ with an average precision in the scaled Hilbert-Schmidt norm squared of \mbox{$\frac{1}{N}||\hat{U}_{\bm \theta} - \hat{U}_{\text{QFT}}||^2_{HS} = 0.00444\pm 0.00138$} and an average final fidelity of $F = 0.9962\pm 0.0014$ between $\hat{U}_{\bm \theta}$ and $\hat{U}_{\text{QFT}}$ acted on the same ensemble of randomly-generated states. The details of how the parameters $\bm \theta$ were acquired can be found in Appendix~\ref{Appendix: QFT} and the code plus parameter values can be found in the GitHub repository~\cite{RVQC_Github}.

\lettersec{Discussion} The RVQC model developed here explicitly demonstrates how one can construct a universal quantum computer out of a set of 1 and 2-qubit UDW time-evolution operators, which uses the relativistic motion combined with the internal dynamics of the qubits to perform computations. We also obtained analytic expressions for the gates, allowing us to directly estimate the effects of the noise from the qubit-field interactions, proving the RVQC could perform meaningful computation in a specific physical regime, which could also be adapted to more general setups. For instance, in curved spacetimes, the tunable parameters would also depend on the geometry of spacetime.

The integration of spacetime geometry into the operation of the RVQC furnishes additional opportunities to employ QML algorithms for learning spacetime properties, representing a unique application of our setting worth further investigation. This makes large-scale trainability concerns such as barren plateaus \cite{cerezo2021cost,holmes2022connecting} less relevant here. Furthermore, promising approaches to circumvent training problems in VQCs do exist~\cite{VQCtraining1, VQCtraining2, VQCtraining3}. 


The key distinctions between the non-relativistic setups commonly found in the literature~\cite{Cerezo_Arrasmith_Babbush_Benjamin_Endo_Fujii_McClean_Mitarai_Yuan_Cincio_2021,Leone_Oliviero_Cincio_Cerezo_2024} and the one presented here are 1) the trainable parameters $\bm \theta$ are prescribed by the spacetime motion of the qubits through the proper time experienced during their evolution; 2) all qubit interactions occur indirectly through each of their local interactions with a relativistic quantum field that is not directly manipulated; and 3) the presence of noise terms due to entanglement between the qubits and the quantum field. 

There are two complementary ways in which one can look at this RVQC setup. On the one hand, it can be seen as a generalization of a familiar spin-spin interaction setup, where the mediating electromagnetic field is explicitly taken into account. In fact, in the simplest case where the qubits are inertial, the trainable parameters can be thought of as the magnitude and direction of an external, tunable magnetic field. On the other hand, this setup is a limit of the full quantum field theoretic description of the interaction between effective qubits, where the quantum degrees of freedom of the mediating field are taken into account, adding noise to the system due to entanglement between the qubits and the field. However, in the current setup, the entanglement present in the field cannot yet be used to implement entangling qubit operations without the need for communication (as is the case in entanglement harvesting protocols~\cite{WhenHarvesting,roleDOF}). Extending the RVQC framework to widely distributed settings where this feature of quantum fields can be harnessed is among our next goals. 

Further building on the prospect of QML in relativistic settings, our construction integrates the properties of relativistic quantum fields into the learning process itself, introducing the idea of learning/computing in environments embedded in spacetime. This raises interesting questions, such as whether one can identify quantum computation in natural phenomena by observing their relativistic motion. Additionally, our work adds new perspectives for establishing large spacetime quantum networks and distributed cryptographic schemes, both of which are at the forefront of quantum technologies. \newline

\begin{acknowledgments}

The authors would like to thank Hendrik Poulsen Nautrup for insightful discussions regarding universality proofs in quantum computing. TRP acknowledges support from the Natural Sciences and Engineering Research Council of Canada (NSERC) via the Vanier Canada Graduate Scholarship. Research at Perimeter Institute is supported in part by the Government of Canada through the Department of Innovation, Science and Industry Canada and by the Province of Ontario through the Ministry of Colleges and Universities. This research was funded in part by the Austrian Science Fund (FWF) [SFB BeyondC F7102, 10.55776/F71]. For open access purposes, the authors have applied a CC BY public copyright license to any author-accepted manuscript version arising from this submission. We gratefully acknowledge support from the European Union (ERC Advanced Grant, QuantAI, No. 101055129). The views and opinions expressed in this article are however those of the author(s) only and do not necessarily reflect those of the European Union or the European Research Council - neither the European Union nor the granting authority can be held responsible for them. 

\end{acknowledgments}

\bibliography{RQML}

\begin{thebibliography}{52}
\bibitem [{foo(note{\natexlab{a}})}]{footnote1}%
  \BibitemOpen
  \href@noop {} {}\bibinfo {howpublished} {{Notice that in this model we are neglecting any local noise experienced by the qubits in our implementation of the local unitaries. This effect could be incorporated by considering the typical $T_1$ and $T_2$ decoherence processes~\cite{BlochT1T2}, and by making sure that the time of implementation of the local unitaries is sufficiently smaller than the $T_1$ and $T_2$ relaxation times}} (\bibinfo {year} {footnote}{\natexlab{a}})\BibitemShut {NoStop}%
\bibitem [{foo(note{\natexlab{b}})}]{footnote2}%
  \BibitemOpen
  \href@noop {} {}\bibinfo {howpublished} {{A quasi-free state is a zero mean Gaussian state. Typical states that satisfy this condition are thermal states and most choices of vacua \cite{Haag}}} (\bibinfo {year} {footnote}{\natexlab{b}})\BibitemShut {NoStop}%
\bibitem [{foo(note{\natexlab{c}})}]{footnote3}%
  \BibitemOpen
  \href@noop {} {}\bibinfo {howpublished} {{This would not be the case if the field was in a highly excited state, or if the qubits were in extreme states of motion while interacting with the field (such as accelerations high enough so that Unruh-like effects would take place~\cite{Unruh-Wald,Takagi,waitUnruh})}} (\bibinfo {year} {footnote}{\natexlab{c}})\BibitemShut {NoStop}%
\bibitem [{foo(note{\natexlab{d}})}]{footnote4}%
  \BibitemOpen
  \href@noop {} {}\bibinfo {howpublished} {{Notice that having $N^2\ll \lambda^{-2} $ does not prevent us from having meaningful entanglement generated by the unitary $\hat{U}_\tc{c}$, provided that $|\lambda^2\Delta_{ij}|\sim 1$. This is a reasonable assumption provided that $T\sim \lambda^{-2} L$, where $T$ is the time duration of the interaction with the field and $L$ is the minimal separation between qubits in the setup}} (\bibinfo {year} {footnote}{\natexlab{d}})\BibitemShut {NoStop}%
\bibitem [{foo(note{\natexlab{e}})}]{footnote5}%
  \BibitemOpen
  \href@noop {} {}\bibinfo {howpublished} {{This assumption corresponds to the statement that the effective coupling range of the qubits is much smaller than their separation, and that their separation is much smaller than the distance traversed by light in a time interval $T$}} (\bibinfo {year} {footnote}{\natexlab{e}})\BibitemShut {NoStop}%
\end{thebibliography}

\appendix

\onecolumngrid

\section{Non-Perturbative Computation for $N$ UDW Qubits}\label{Appendix:Calculations}

Our channel $\mathcal{E}$ is built by assuming an interaction with the qubits and the field. We have $N$ qubits that undergo timelike trajectories $\mf z_i(\tau)$. The interaction with the field $\hat{\phi}(\mf x)$ is described by the interaction Hamiltonian density
\begin{equation}\label{eq:hIappendix}
    \hat{h}_I(\mf x) = \lambda \sum_{i=1}^N\hat{\mu}_i \Lambda_i(\mf x) \hat{\phi}(\mf x).
\end{equation}
In this appendix we derive the final density operator of the $N$-qubit state after its interaction with the field in a quasifree state. In what follows, we define the smeared field operators
\begin{align}
    \hat{\phi}(\Lambda_i) = \int \dd V \Lambda_i(\mf x) \hat{\phi}(\mf x),
\end{align}
where $\dd V$ is the spacetime volume element. We denote the retarded and advanced Green's functions associated with the field's equation of motion by $G_R(\mf x,\mf x')$ and $G_A(\mf x, \mf x')$, respectively. We can then write the covariant canonical commutation relations for the field as
\begin{equation} \label{Equation:CovCanCom}
    [\hat{\phi}(f), \hat{\phi}(g)] = \ii E(f,g),
\end{equation}
where $E(\mf x, \mf x') = G_R(\mf x, \mf x') - G_A(\mf x, \mf x')$ is the causal propagator and we use the convention that bi-scalars act on two spacetime functions according to
\begin{equation}
    E(f,g) = \int \dd V \dd V' f(\mf x) g(\mf x') E(\mf x, \mf x').
\end{equation}
In particular, the smeared field's two-point function in a state $\hat{\rho}_\phi$ can be written as
\begin{equation}
    W(f,g) = \int \dd V \dd V' f(\mf x) g(\mf x') W(\mf x, \mf x'),
\end{equation}
where $W(\mf x, \mf x') = \tr(\hat{\phi}(\mf x) \hat{\phi}(\mf x')\hat{\rho}_\omega)$ is the so-called Wightman function.

The Magnus expansion can be used to compute the time evolution operator \cite{Magnus_2009} associated with the Hamiltonian denstiy~\eqref{eq:hIappendix}. We find
\begin{equation}
    \hat{U}_I = e^{\hat{\Theta}_1+\hat{\Theta}_2},\label{eq:UImag}
\end{equation}
where
\begin{align}
    \hat{\Theta}_1 &= - \ii \int \dd V \hat{h}_I(\mf x) = - \ii \lambda \sum_{i=1}^N\hat{\mu}_i \hat{\phi}(\Lambda_i),\\
    \hat{\Theta}_2 &= - \frac{1}{2} \int \dd V \dd V' \theta(t-t')[\hat{h}_I(\mf x), \hat{h}_I(\mf x')] \label{Equation:TH2}\\&= - \frac{\lambda^2}{2}\int \dd V \dd V' \theta(t-t')[\hat{\phi}(\mf x), \hat{\phi}(\mf x')]\Big(\sum_{i=1}^N\hat{\mu}_i^2 \Lambda_i(\mf x)\Lambda_i(\mf x') + \sum_{{\substack{{}_{i,j=1}\\{}_{i\neq j}}}}^N\hat{\mu}_i\hat{\mu}_j \Lambda_i(\mf x)\Lambda_j(\mf x')\Big)\nonumber\\
    &= - \frac{\ii \lambda^2}{2}\Big(\sum_{i=1}^N\hat{\mu}_i^2 G_R(\Lambda_i,\Lambda_i)+\sum_{{\substack{{}_{i,j=1}\\{}_{i\neq j}}}}^N\hat{\mu}_i\hat{\mu}_j G_R(\Lambda_i,\Lambda_j)\Big),\nonumber\\
    &= - \frac{\ii \lambda^2}{2}\Big(\sum_{i=1}^N\hat{\mu}_i^2 G_R(\Lambda_i,\Lambda_i)+\sum_{{\substack{{}_{i,j=1}\\{}_{i< j}}}}^N\hat{\mu}_i\hat{\mu}_j \Delta(\Lambda_i,\Lambda_j)\Big),\nonumber
\end{align}
where we used $\theta(t-t')[\hat{\phi}(\mf x),\hat{\phi}(\mf x')] = \ii G_R(\mf x, \mf x')$ and introduced $\Delta_{ij} = \Delta(\Lambda_i, \Lambda_j) = G_R(\Lambda_i,\Lambda_j)+G_R(\Lambda_j,\Lambda_i)$ as the symmetric propagator. Defining $\mathcal{G}_i = G_R(\Lambda_i,\Lambda_i)$, Eq.~\eqref{Equation:TH2} allows us to write
\begin{equation}
    \hat{\Theta}_2 = - \frac{\ii \lambda^2}{2}\Big(\sum_{i=1}^N\mathcal{G}_i\hat{\mu}_i^2 +\sum_{{\substack{{}_{i,j=1}\\{}_{i< j}}}}^N\Delta_{ij}\hat{\mu}_i\hat{\mu}_j \Big)\nonumber.
\end{equation}
A consequence of the covariant canonical commutation relations Eq. \eqref{Equation:CovCanCom} and the definition of $\hat{h}_I(\mf x)$ Eq. \eqref{eq:hIappendix} is that $[[\hat{h}_I(\mf x), \hat{h}_I(\mf x')],\hat{h}_I(\mf x'')] = 0$, which implies that only the first two terms $\hat{\Theta}_1$ and $\hat{\Theta}_2$ of the Magnus expansion as shown in Eq.~\eqref{eq:UImag} are non-zero, so that the unitary time evolution operator reads
\begin{equation}\label{Equation:UI2g}
    \hat{U}_I = \exp(- \ii \lambda \sum_{i=1}^N\hat{\mu}_i \hat{\phi}(\Lambda_i))\exp(- \frac{\ii \lambda^2}{2}\sum_{i=1}^N\mathcal{G}_i\hat{\mu}_i^2)\exp( -\frac{\ii \lambda^2}{2}\sum_{{\substack{{}_{i,j=1}\\{}_{i< j}}}}^N\Delta_{ij}\hat{\mu}_i\hat{\mu}_j ).
\end{equation}
One can also use the Baker-Campbell-Hausdorff formula in order to factor $\hat{U}_I$ so that each local operator in the qubits and fields acts as an individual unitary. This calculation is not useful for reaching the final state of the qubits, but might be insightful for causality analysis. We skip this for the time being.

For simplicity, we assume that $\hat{\mu}_i^2 = \openone$, so that terms depending on $\mathcal{G}_i$ only give global phases and can be neglected in the computation of $\hat{U}_I$. We also factor
\begin{equation}
    \hat{U}_I = \hat{U}_\phi\hat{U}_\tc{c},
\end{equation}
where we define
\begin{align}
    \hat{U}_\phi &= \exp(- \ii \lambda \sum_{i=1}^N\hat{\mu}_i \hat{\phi}(\Lambda_i)),\\
    \hat{U}_\tc{c} &=\exp( -\frac{\ii \lambda^2}{2}\sum_{{\substack{{}_{i,j=1}\\{}_{i< j}}}}^N\Delta_{ij}\hat{\mu}_i\hat{\mu}_j ).
\end{align}
Notice that only the unitary $\hat{U}_\phi$ depends on the field, and that $[\hat{U}_\phi,\hat{U}_\tc{c}] = 0$. Under the assumption that the initial state for the qubit-field system is $\hat{\rho}_0\otimes \hat{\rho}_\omega$, where $\hat{\rho}_0$ is the initial state of the $N$ qubit system and $\hat{\rho}_\omega$ is a quasi-free state of the field, we can then write the final qubits state as 
\begin{equation}
    \hat{\rho}_f = \tr_\phi(\hat{U}_I (\hat{\rho}_0\otimes \hat{\rho}_\omega)\hat{U}_I^\dagger) = \hat{U}_\tc{c}\tr_\phi(\hat{U}_\phi(\hat{\rho}_0\otimes \hat{\rho}_\omega)\hat{U}_\phi^\dagger)\hat{U}_\tc{c}^\dagger.
\end{equation}
Let us first compute 
\begin{align}
    \hat{U}_\phi (\hat{\rho}_0\otimes \hat{\rho}_\omega) \hat{U}_\phi^\dagger &= \exp(- \ii \lambda \sum_{i=1}^N\hat{\mu}_i \hat{\phi}(\Lambda_i))(\hat{\rho}_{0}\otimes\hat{\rho}_\omega) \exp(- \ii \lambda \sum_{j=1}^N\hat{\mu}_j \hat{\phi}(\Lambda_j))\nonumber\\
    &=\!\!\!\!\sum_{\mu_1,\mu_1' = \pm}\cdots \sum_{\mu_N,\mu_N' = \pm} \!\!e^{- \ii \lambda\hat{\phi}(\sum_{i=1}^N{\mu}_i\Lambda_i)}\hat{\rho}_\omega e^{\ii \lambda\hat{\phi}(\sum_{j=1}^N{\mu}_j'\Lambda_j)}\bra{\mu_1...\mu_N}\hat{\rho}_0\ket{\mu_1'...\mu_N'}  \ket{\mu_1...\mu_N}\!\!\bra{\mu_1'...\mu_N'}.
\end{align}
We define the qubit channel $\mathcal{E}_\phi(\hat{\rho}_0) \coloneqq \tr_\phi(\hat{U}_\phi (\hat{\rho}_0\otimes \hat{\rho}_\omega) \hat{U}_\phi^\dagger)$, so that
\begin{align} \label{Equation:sigf}
    \mathcal{E}_\phi(\hat{\rho}_0)
     &=\!\!\!\!\sum_{\mu_1,\mu_1' = \pm}\cdots \sum_{\mu_N,\mu_N' = \pm} \!\!\omega\left( e^{\ii \lambda\hat{\phi}(\sum_{j=1}^N{\mu}_j'\Lambda_j)}e^{- \ii \lambda\hat{\phi}(\sum_{i=1}^N{\mu}_i\Lambda_i)}\right)\bra{\mu_1...\mu_N}\hat{\rho}_0\ket{\mu_1'...\mu_N'}  \ket{\mu_1...\mu_N}\!\!\bra{\mu_1'...\mu_N'}  \\  &=\!\!\!\!\sum_{\mu_1,\mu_1' = \pm}\cdots \sum_{\mu_N,\mu_N' = \pm} \!\!e^{-\frac{\ii \lambda^2}{2} E(\sum_{j=1}^N{\mu}_j'\Lambda_j,\sum_{i=1}^N{\mu}_i\Lambda_i) - \frac{\lambda^2}{2} \left|\left|\sum_{i=1}^N (\mu_i - \mu'_{i})\Lambda_i\right|\right|^2 }\bra{\mu_1...\mu_N}\hat{\rho}_0\ket{\mu_1'...\mu_N'}  \ket{\mu_1...\mu_N}\!\!\bra{\mu_1'...\mu_N'},\nonumber
\end{align}
where we used
\begin{equation}
    \omega\left(e^{\ii \lambda \hat{\phi}(f)}e^{\ii \lambda \hat{\phi}(g)}\right) = e^{\frac{\ii \lambda^2}{2}E(f,g) - \frac{\lambda^2}{2} W(f+g,f+g)},
\end{equation}
with $||f||^2 = W(f,f)$.
\begin{equation}
    \hat{\rho}_f = \mathcal{E}(\hat{\rho}_0) = \hat{U}_\tc{c} \mathcal{E}_\phi(\hat{\rho}_0)\hat{U}_\tc{c}^\dagger= \exp( -\frac{\ii \lambda^2}{2}\sum_{{\substack{{}_{i,j=1}\\{}_{i< j}}}}^N\Delta_{ij}\hat{\mu}_i\hat{\mu}_j )\mathcal{E}_\phi(\hat{\rho}_0) \exp(\frac{\ii \lambda^2}{2}\sum_{{\substack{{}_{i,j=1}\\{}_{i< j}}}}^N\Delta_{ij}\hat{\mu}_i\hat{\mu}_j ).\label{Equation:sab}
\end{equation} 
The final channel is then a combination of multiple commuting entangling unitaries between each qubit and the added quantum channel $\mathcal{E}_\phi$ obtained when one traces over the field. Under the assumption that $E(\Lambda_i,\Lambda_j) = 0$ (which occurs for instance when the interaction regions are only shifted in space), $\mathcal{E}_\phi$ acts merely as a decohering channel. 

\section{Lower Bounds on the Fidelity Between $\mathcal{E}_\phi$ and $\hat{U}_\tc{c}$} \label{Appendix:Fidelity}

We can also estimate to which extent the channel $\mathcal{E}$ can be approximated by the unitary $\hat{U}_\tc{c}$. To do so, we can compute the fidelity between the resulting states $\hat{\rho}_f, \hat{\rho}_c$ due to the time evolution of an arbitrary initial state $\hat{\rho}_0$ under the channel $\mathcal E$ and $\hat{U}_c$, respectively \cite{nielsen2010quantum}.  The fidelity is given by the expression
\begin{align} \label{Equation:Fidelity}
F(\hat{\rho}_f, \hat{\rho}_c) = F(\mathcal{E}(\hat{\rho}_0), \hat{U}_c\hat{\rho}_0\hat{U}^{\dagger}_c) = \left(\text{Tr}\left(\sqrt{\sqrt{\hat{\rho}_f}\hat{\rho}_c\sqrt{\hat{\rho}_f}}\right)\right)^2\, .
\end{align}
Since Eq. \ref{Equation:Fidelity} is quite difficult to compute for arbitrary states, we can first lower bound the fidelity using the trace inequality $0 \leq\text{Tr}\left(A^2\right) \leq \left(\text{Tr}\left(A\right)\right)^2 $, for a positive semi-definite operator $A$~\cite{Yang_Feng_2002}, as:
\begin{align} \label{Equation:FidelityBound}
0\leq \text{Tr}\left(\sqrt{\hat{\rho}_f}\hat{\rho}_c\sqrt{\hat{\rho}_f}\right) = \text{Tr}\left(\hat{\rho}_f\hat{\rho}_c\right) = \text{Tr}\left(\hat{U}_c\mathcal{E}_\phi(\hat{\rho}_0)\hat{U}^{\dagger}_c\hat{U}_c\hat{\rho}_0\hat{U}^{\dagger}_c\right) = \text{Tr}\left(\mathcal{E}_\phi(\hat{\rho}_0)\hat{\rho}_0\right)\leq F(\hat{\rho}_f, \hat{\rho}_c)
\end{align}
where we have used the cyclic property of the trace twice and applied the definition Eq. \eqref{Equation:sab} of $\mathcal{E}$. We can explicitly express Eq. \eqref{Equation:FidelityBound}
 (again under the assumption that $E(\Lambda_i,\Lambda_j) = 0$) as
\begin{align}
\text{Tr}\left(\mathcal{E}_\phi(\hat{\rho}_0)\hat{\rho}_0\right)
    &=\!\!\!\!\sum_{\mu_1,\mu_1' = \pm}\cdots \sum_{\mu_N,\mu_N' = \pm} \!\!e^{ - \frac{\lambda^2}{2} \left|\left|\sum_{i=1}^N (\mu_i - \mu'_{i})\Lambda_i\right|\right|^2 }\nonumber \\[1ex] 
&\qquad \qquad \qquad \times\text{Tr}\Big(\bra{\mu_1...\mu_N}\hat{\rho}_0\ket{\mu_1'...\mu_N'} \ket{\mu_1...\mu_N}\!\!\bra{\mu_1'...\mu_N'}\hat{\rho}_0\Big),
 \label{Equation:BigOverlap}
\end{align}

Now let us study the terms $e^{ - \frac{\lambda^2}{2} \left|\left|\sum_{i=1}^N (\mu_i - \mu'_{i})\Lambda_i\right|\right|^2}$ in more detail. We can expand the argument and bound it as follows
\begin{align}
    \left|\left|\sum_{i=1}^N (\mu_i - \mu'_{i})\Lambda_i\right|\right|^2 &=  W\left(\sum_{i=1}^N(\mu_i - \mu'_{i})\Lambda_i,\sum_{j=1}^N(\mu_j - \mu'_{j})\Lambda_j\right) \\&= \sum_{i,j=1}^N (\mu_i - \mu'_{i})(\mu_j - \mu'_{j})W\left(\Lambda_i,\Lambda_j\right)\leq \mathcal W \sum_{i,j=1}^N\left| (\mu_i - \mu'_{i})(\mu_j - \mu'_{j})\right|,
\end{align}
where we defined
\begin{equation}
    \mathcal{W} = \max_{i,j}(|W(\Lambda_i,\Lambda_j)|) = \max_{i}(|W(\Lambda_i,\Lambda_i)|).
\end{equation}
Notice that the sum of $\mu_i + \mu_i'$ can only be $\pm2$ or $0$, as these only take the values of $1$ and $-1$. The maximum value that can be achieved by the sums above then happens when $\mu_i = -\mu_i'$, leading to an upper bound of 
\begin{equation}
    \left|\left|\sum_{i=1}^N (\mu_i - \mu'_{i})\Lambda_i\right|\right|^2 \leq 4 N^2 \mathcal{W}.
\end{equation}
Using the bound above in Eq.~\eqref{Equation:BigOverlap}, and noticing that
\begin{align}
\!\!\!\!\sum_{\mu_1,\mu_1' = \pm}\cdots \sum_{\mu_N,\mu_N' = \pm} \!\!\text{Tr}\Big(\bra{\mu_1...\mu_N}\hat{\rho}_0\ket{\mu_1'...\mu_N'} \ket{\mu_1...\mu_N}\!\!\bra{\mu_1'...\mu_N'}\hat{\rho}_0\Big) = \text{Tr}(\hat{\rho}_0^2),
\end{align}
we find the following lower bound for the fidelity corresponding to an arbitrary quantum state $\hat{\rho}_0$:
\begin{equation}
 e^{-2\lambda^2N^2 \mathcal{W}}\text{Tr}\left(\hat{\rho}_0^2\right)\leq\text{Tr}\left(\mathcal{E}_\phi(\hat{\rho}_0)\hat{\rho}_0\right)\leq F(\hat{\rho}_f, \hat{\rho}_c), \label{Equation:FidelityExponentialBound}
\end{equation}
where we used $\text{Tr}\left(\hat{\rho}_0^2\right) \leq \text{Tr}\left(\hat{\rho}_0\right) = 1$.

At first one might lament that a lower bound which is a Gaussian in the number of qubits is too loose to be useful as it is effectively zero for almost all interesting circuits, which essentially means it is not a lower bound because the fidelity is already bounded below by 0 (since it is a probability). However, the effective UDW treatment for the qubits requires the coupling constant $\lambda$ to be sufficiently small. For instance, if $\lambda$ were of the order of $10^{-4}$, $e^{-2N^2\lambda^2\mathcal W} \approx 1$ for a reasonably large number of qubits. It is also common to consider pure initial states, for which the purity factor evaluates to 1. Therefore, this lower bound is actually very tight for most applications and the channel $\mathcal E$ is close to unitary.

\section{Closed-Form Expressions for the Explicit Setup}\label{Appendix:ClosedForm}

We consider the interactions to be defined by the spacetime smearing functions
\begin{equation}
    \Lambda_i(\mf x) = e^{- \frac{(t - t_\tc{c})^2}{2T^2}}\frac{e^{- \frac{|\bm x - \bm x_i|^2}{2\sigma^2}}}{(2\pi \sigma^2)^\frac{3}{2}} = \Lambda(t-t_\tc{c}, \bm x - \bm x_i),
\end{equation}
so that the interactions of the qubits with the field is centered at $t = t_{\tc{c}}$, where the spacetime smearing functions are centered and can be well approximated to take a time of $14T$, which contains $99.99999999974\%$ of the support of the spacetime smearing function when centered at $t = t_\tc{c}$. Neglecting the effect of the tails outside the interval $-7T < t - t_\tc{c} < 7T$, we can compute the expressions for $W(\Lambda_i,\Lambda_j)$, $E(\Lambda_i,\Lambda_j)$ and $\Delta(\Lambda_i,\Lambda_j)$ analytically in terms of the parameters $T$, $\sigma$ and $L_{ij} = |\bm x_i - \bm x_j|$ in Minkowski space. We find that $E(\Lambda_i,\Lambda_j) = 0$,
\begin{align}
    W(\Lambda_i,\Lambda_i) &= \frac{1}{4 \pi\alpha^2}\\
    W(\Lambda_i,\Lambda_j) &= \frac{T}{4 \sqrt{\pi} \alpha L_{ij} }e^{- \frac{L_{ij}^2}{4 \alpha^2 T^2}}\text{erfi}\left(\frac{L_{ij}}{2\alpha T}\right)\\
    \Delta(\Lambda_i,\Lambda_j) &= -\frac{T}{2\sqrt{\pi} \alpha L_{ij}}e^{-\frac{L_{ij}^2}{4\alpha^2 T^2 }}\text{erf}\left(\frac{L_{ij}}{2 \alpha\sigma}\right),
\end{align}
where $\alpha = \sqrt{1 + \sigma^2/T^2}$. Also notice that on the limit of $T\gg L_{ij}$, we have $W(\Lambda_i,\Lambda_j)\approx W(\Lambda_i,\Lambda_i)$. The terms $\Delta_{ij}$ behave proportionally to $T/L_{ij}$.

\section{Proof of Universality}
\label{Appendix:Universality}

In this appendix, we will argue that our single-qubit rotations, together with a transpiled version of the fixed entangling gate 
\begin{align}
    \hat{U}_\tc{c} &=\exp( -\frac{\ii \lambda^2}{2}\sum_{{\substack{{}_{i,j=1}\\{}_{i< j}}}}^N\Delta_{ij}Z_i Z_j ),
\end{align}
are universal for quantum computation. Here, we write $X \equiv \hat{\sigma}_X$, $Y \equiv \hat{\sigma}_Y$, $Z \equiv \hat{\sigma}_Z$. Since we already have arbitrary local $\mathrm{SU}(2)$-rotations, we only have to show that we can transpile $\hat{U}_c$ and local rotations into two-qubit entangling gates on arbitrary qubit pairs in order to have a universal gate set. Given this, considering each qubit pair as its own quantum system, we can transpile $\mathrm{CNOT}$ gates for each qubit pair, and combining these $\mathrm{CNOT}$-gates, we get a universal gate set. 

Without loss of generality, assume that the qubits are labelled such that we want to get an entangling gate on qubits $N-1$ and $N$. First, we take advantage of the fact that $X_j \cdot e^{\ii Z_1 \otimes Z_2} \cdot X_j = e^{- \ii Z_1 \otimes Z_2}$ with $j \in \{1,2\}$ to find 
\begin{align}
    \hat{U}^{(1)} := (X_{N-1} \otimes X_{N})\hat{U}_c (X_{N-1} \otimes X_{N}) \hat{U}_c = \exp( -\ii \lambda^2\sum_{{\substack{{}_{i,j=1}\\{}_{i< j}}}}^{N-2}\Delta_{ij}Z_i Z_j ) \exp( - \ii \lambda^2\Delta_{N-1, N}Z_{N-1} Z_{N} ).
\end{align}
Here, the last exponential looks like the expression we want, and we will abbreviate $\hat{U}_{\mathrm{pair}} := \exp( -\ii \lambda^2\Delta_{N-1, N}Z_{N-1} Z_{N} )$. Furthermore, we recognize that the first exponential is similar to $U_c$, but only acts on $N-2$ qubits. So far, we have consumed two layers of the VQC.

For the next step, we split the qubit set $\{1,2,\dots , N-2\}$ into two sets which we call $\mathcal I^{(1)}_X$ and $\mathcal I^{(1)}_{\mathbb 1}$. Here, each set should contain roughly half of the qubits, which we fix by demanding $|\mathcal I^{(1)}_X| = \lfloor\frac{N-2}{2}\rfloor$.

Now, we define the next iteration step as 
\begin{align}
    \hat{U}^{(2)} := \bigotimes_{k \in \mathcal{I}^{(1)}_X} X_k \hat{U}^{(1)} \bigotimes_{m \in \mathcal{I}^{(1)}_X} X_m \hat{U}^{(1)} = \exp( -2 \ii \lambda^2 \sum_{ i < j \in \mathcal{I}^{(1)}_X}\Delta_{ij}Z_i Z_j ) \exp( - 2\ii \lambda^2\sum_{ i < j \in \mathcal{I}^{(1)}_{\mathbb 1}}\Delta_{ij}Z_i Z_j ) \hat{U}_{\mathrm{pair}}^2
\end{align}
We have consumed two $U^{(1)}$, each of which costs two VQC layers, bringing us to a cost of $4 = 2^2$ layers.

Next, we split  both $\mathcal{I}^{(1)}_X$ and $\mathcal{I}^{(1)}_{\mathbb 1}$ into halves, introducing the sets $\mathcal{I}^{(2)}_{\mathbb 1, \mathbb 1}$, $\mathcal{I}^{(2)}_{X, \mathbb 1}$, $\mathcal{I}^{(2)}_{\mathbb 1, X}$, and $\mathcal{I}^{(2)}_{X, X}$. For the sizes, we demand that $|\mathcal{I}^{(2)}_{\mathbb 1, X}| = \lfloor \frac{|\mathcal{I}^{(1)}_{\mathbb 1} |}{2} \rfloor$ and $|\mathcal{I}^{(2)}_{X, X}| = \lfloor \frac{|\mathcal{I}^{(1)}_{X} |}{2} \rfloor$. Note, we use the convention that $\sum_{i< j \in \varnothing} \Delta_{ij} Z_i Z_j = 0$ and $\sum_{i< j \in \{ k \}} \Delta_{ij} Z_i Z_j = 0$, i.e. empty index sets and one-element index sets just multiply the unitary with the identity. Furthermore, we use the convention that $\otimes_{k \in \varnothing} X_k = \mathbb 1$.

Now, the next iteration step gives
\begin{align}
    \hat{U}^{(3)} := \bigotimes_{k \in \mathcal{I}^{(2)}_{\mathbb 1, X} \cup \mathcal{I}^{(2)}_{X, X}} X_k \hat{U}^{(2)}  \bigotimes_{m \in \mathcal{I}^{(2)}_{\mathbb 1, X} \cup \mathcal{I}^{(2)}_{X, X}} X_m  \hat{U}^{(2)} = \prod_{J_1, J_2 \in \{\mathbb 1, X\}} \exp(-2^2 \ii \lambda^2 \sum_{i < j \in \mathcal{I}_{J_1, J_2}^{(2)}} \Delta_{i,j} Z_i Z_j) \hat{U}_{\mathrm{pair}}^{2^2} 
\end{align}

We are in the position to give the general form for the $k^{\text{th}}$ iteration. For $J_1,\dots, J_{k-1} \in \{\mathbb 1 , X\}$, we split the index sets $\mathcal I^{(k-1)}_{J_1, \dots J_{k-1}}$ into two halves, obtaining $\mathcal I^{(k)}_{J_1, \dots J_{k-1}, \mathbb 1}$ and $\mathcal I^{(k)}_{J_1, \dots J_{k-1} , X}$. Again, the specific demand for the sizes is $|\mathcal I^{(k)}_{J_1, \dots J_{k-1} , X}| = \lfloor \frac{ | \mathcal I^{(k-1)}_{J_1, \dots J_{k-1}} |}{2}\rfloor$. Then, the next iteration step is defined as
\begin{align}
    \hat{U}^{(k+1)} := &\bigotimes_{\substack{J_1,\dots J_{k-1} \in \{ \mathbb 1 , X\}, \\ m \in \mathcal I^{(k)}_{J_1, \dots , J_{k-1}, X}} } X_m \hat{U}^{(k)} \bigotimes_{\substack{\tilde{J}_1,\dots \tilde{J}_{k-1} \in \{ \mathbb 1 , X\}, \\ \tilde{m} \in \mathcal I^{(k)}_{\tilde{J}_1, \dots , \tilde{J}_{k-1}, X}} } X_{\tilde{m}} \hat{U}^{(k)} \nonumber \\ = & \prod_{J_1, \dots , J_k \in \{\mathbb 1, X\}} \exp(-2^{k} \ii \lambda^2 \sum_{i < j \in \mathcal{I}_{J_1, \dots , J_k}^{(k)}} \Delta_{i,j} Z_i Z_j) \hat{U}_{\mathrm{pair}}^{2^{k}} 
\end{align}
We have consumed two $\hat{U}^{(k)}$, each of which costs $2^k$ VQC layers, giving a combined cost of $2^{k+1}$ VQC layers. 

Let us analyze the general form of $\hat{U}^{(k+1)}$ in more detail. First of all, we recognize that interactions are only present for $i,j$ in the same index set $\mathcal I_{J_1,\dots , J_k}^{(k)}$ (and $\{N-1, N\}$, of course). 

Furthermore, the condition $|\mathcal{I}^{(k)}_{J_1, \dots J_{k-1} , X}| = \lfloor \frac{ |\mathcal{I}^{(k-1)}_{J_1, \dots J_{k-1}} |}{2}\rfloor$ expresses that the index sets are exponentially decreasing in size. We will use this observation to show that we can terminate the iteration scheme after a number of steps no larger than $\lceil \log_2 N \rceil + 3$.

For this purpose, we will first prove the following lemma: 
\begin{align}
\max_{J_1,\dots, J_k} |\mathcal{I}^{(k)}_{J_1,\dots , J_k}| \le \max\left\{ \frac{N}{2^k} +2 -\frac{1}{2^{k-1}} \ , \ 1 \right\}
\end{align}

For $k=1$, the first entry in the maximum evaluates to $\frac{N}{2} + 1$. This is a true statement, because we start by splitting $N-2$ qubits into two sets no larger than $\lceil\frac{N-2}{2} \rceil =  \lceil\frac{N}{2} -1 \rceil = \lceil\frac{N}{2} \rceil -1 < \frac{N}{2} + 1$.

Assume now that the statement is true for $k$, and let us show that it also holds for $k+1$. If $1$ is the maximum on the right hand side, we have finished because there are no more qubits left entangled (except for $N-1$ and $N$, of course). So let us assume now that all $|\mathcal I^{(k)}_{\dots}| \le \frac{N}{2^k} + 2 - \frac{1}{2^{k-1}}$. The worst cardinality that the iteration scheme might produce is therefore $\left\lceil\frac{\frac{N}{2^k} + 2 - \frac{1}{2^{k-1}} }{2}\right\rceil = \left\lceil\frac{N}{2^{k+1}} + 1 -\frac{1}{2^{k}}\right\rceil < \frac{N}{2^{k+1}} + 2 -\frac{1}{2^k}$, finishing the induction proof.

Therefore, we find that the $|\mathcal{I}^{(k)}_{J_1,\dots , J_k}|$ are either upper-bounded by $1$, or by $\frac{N}{2^k} + 2$. If we choose $k \equiv \lceil \log_2 N \rceil$, we find that $|\mathcal{I}^{(k)}_{J_1,\dots , J_k}| \le 3$. Iterating the scheme two more times then achieves that all $|\mathcal{I}^{(k+2)}_{J_1,\dots , J_{k+2}}| \le 1$.

Therefore, together with our conventions on empty and one-element index sets, we find that 
\begin{align}
    \hat{U}^{(\lceil \log_2 (N)\rceil + 3)} = \hat{U}_{\mathrm{pair}}^{ 2^{\lceil \log_2 (N)\rceil +2}} \label{Equation:OnlyQubitPairLeft}
\end{align}
has a cost of $\mathcal O (2^{\lceil \log_2 (N)\rceil}) = \mathcal O(N)$, i.e. it is linear in the number of qubits. In other words, the exponential cost in the number of iterations is compensated by the fact that we only need logarithmically many iterations. {Importantly, for Equation~\eqref{Equation:OnlyQubitPairLeft} to define an entangling unitary acting on the pair, we require that $2^{\lceil \log_2(N) \rceil + 2} \lambda^2 \Delta_{ij}$ is bounded away from multiples of $\pi$. That is, the values of $\Delta_{ij}$ have to be sufficiently inhomogeneous. This is not a limiting constraint, for instance, using the setup where the detectors are inertial during their interaction (see Appendix~\ref{Appendix:ClosedForm}), it can be achieved by perturbing the separations between the qubits $L_{ij}$.}

As a technical side-note, instead of using all the ceiling and flooring operations, one can introduce $2^{\lceil \log_2 (N)\rceil} + 2 - N$ fictitious qubits, and formally set  $\Delta_{ij} = 0$ whenever $i$ or $j$ is one of those extra fictitious qubits. The $+2$ is because of the qubit pair we actually want to entangle. Then, $|\mathcal{I}^{(k)}_{J_1, \dots J_{k-1} , X}| = \lfloor \frac{ |\mathcal{I}^{(k-1)}_{J_1, \dots J_{k-1}} |}{2}\rfloor = \frac{ |\mathcal{I}^{(k-1)}_{J_1, \dots J_{k-1}} |}{2}$ without rounding, except if we are already at $|\mathcal{I}^{(k-1)}_{J_1, \dots J_{k-1}} | \in \{0,1\}$, the index sets that have no entangling operations.

The other way around, we can also use this construction to remove entangling layers completely: For $m \in \mathbb N$ such that $2^{\lceil \log_2(N) \rceil + 2} m \lambda^2 \Delta_{N-1, N}$ is very close to a multiple of $2\pi$ (such $m$ exists, as can be seen by approximating $\pi$ with rational numbers), we find $\Big(\hat{U}_{\mathrm{pair}}^{2^{\lceil \log_2 (N)\rceil +2}} \Big)^m \approx \mathbb 1 $. Alternatively, one just eliminates $\hat{U}_{\mathrm{pair}}$ using $X_N \hat{U}^{\lceil \log_2(N) \rceil +2}_{\mathrm{pair}} X_N \hat{U}^{\lceil \log_2(N) \rceil +2}_{\mathrm{pair}}$ analogously to before.

\section{Quantum Fourier Transform}
\label{Appendix: QFT}

Here we will discuss the details on the implementation of the quantum Fourier transform, going into detail on how the parameters $\bm \theta$ mentioned at the explicit setup were obtained. 

The procedure begins by setting the hyperparameters from the end of the explicit setup section and then uniformly randomly initializing the parameters $\bm \theta = \{\theta_i, \vartheta_i, \varphi_i\}_i^N$, on the intervals $[0, 10\pi), [0, 2\pi), [0, 2\pi)$, for each layer in the relativistic VQC. The task is then to minimize a loss function $\mathcal L(\bm \theta)$ that encodes the details of the quantum Fourier transform $\hat{U}_{\text{QFT}}$ and thereby obtain updates for the parameters that yield increasingly better approximations of $\hat{U}_{\text{QFT}}$ (the Qiskit implementation) using our model $\hat{U}_{\bm \theta}$. Since our platform's architecture is already in the form of a VQC, we only need to specify the loss function and optimization algorithm to get started.

The quantum Fourier transform learning task with the inefficient loss function given by the scaled Hilbert-Schmidt norm squared
\begin{align}
\mathcal L(\bm \theta) = \frac{1}{N}\left|\left|\hat{U}_{\bm \theta} - \hat{U}_{\text{QFT}}\right|\right|_{HS}^2 = \frac{1}{N}\sum_{ij}\left |u^{\bm \theta}_{ij} - u^{\text{QFT}}_{ij}\right|^2 \,,
\end{align}
was chosen for illustration purposes such that a data science background is not required.

\begin{figure}[h!]
    \includegraphics[width = 0.8 \linewidth]{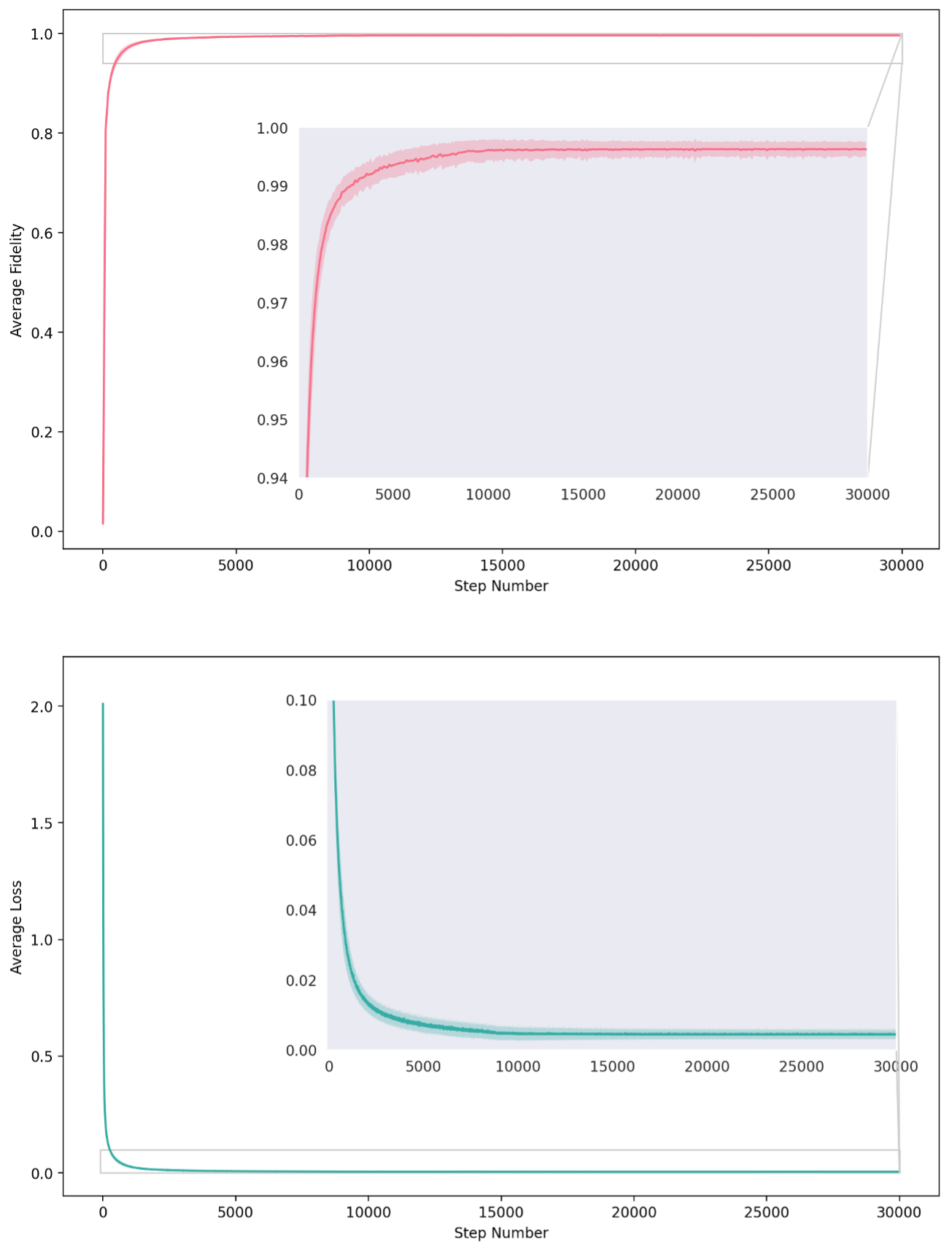}
    \caption{The average fidelity (top) and loss (bottom) over 20 different training runs. One can see that the model quickly reaches a low/high loss/fidelity but takes some time to converge to the loss tolerance value, which is typical behaviour in similarly-sized QML problems. Each training run took a different total number of steps to converge, the lowest being 4717 and the highest being 30000. To do the plot averaging, we padded the values of all runs smaller than the longest one with their final converged value, which is akin to freezing the learning rate after convergence. Note that the top figure displays the average fidelity calculated every 100 training steps. The lighter, shaded regions are ± the standard deviation of the average fidelity/loss, which is calculated over the 20 different runs at the same time step.}
    \label{Figure:FidelityLoss}
\end{figure}

We also use the fidelity between our model $\hat{U}_{\bm \theta}|\psi\rangle$ and the quantum Fourier transform $\hat{U}_{\text{QFT}}|\psi\rangle$, with $|\psi\rangle$ representing the 64 normalized test states used in the calculation, as a measure of how well our model captures the input-output behaviour of the quantum Fourier transform circuit. The coefficients of these test states were initialized from a complex random (standard) normal distribution and the computation of the fidelity only occurred every 100 training steps, as is specified in Figure \ref{Figure:FidelityLoss}. Note that the fidelity is merely an additional measure we employ and is not included as part of the loss function nor does it have any impact on the training process.

The optimization algorithm we use to update the parameters $\bm \theta$ is the Pytorch implementation of the Adam algorithm with the default settings and an initial learning rate of 0.01. For the training, we set a tolerance value of 0.004 for the loss function and a maximum step number of 30000, after which the training period terminates. We also perform the training for 20 different model parameter initializations to mitigate initialization biases and take an average of these runs for the loss and fidelity, as shown in Figure \ref{Figure:FidelityLoss}. Note that the 2nd and 16th runs did not reach the tolerance value within the maximum allowed number of iterations, which we suspect is due to the increased likelihood of encountering bad local minima when using uniform random distributions to initialize the trainable parameters \cite{anschuetz2022quantum,anschuetz2024unified}. However, these runs still achieved an outstanding average final fidelity of $99.16\%$ and $99.26\%$, respectively, despite not reaching the final loss tolerance. 

\end{document}